\documentclass{aa}
\usepackage{graphicx}
\usepackage{txfonts}

\begin{document}

\title{Noise power spectrum estimation and fast map making for CMB experiments}

   \subtitle{}

\author{ 
A.~Amblard \inst{1,3} \and J.--Ch.~Hamilton \inst{2,3,4} }

   \offprints{amblard@in2p3.fr}
   \mail{}
\institute{
University of California, Department of Astronomy, 601 Campbell Hall,
Berkeley, CA 94720-3411, U.S.A.\and Laboratoire de Physique
Nucl\'eaire et de Hautes Energies, 4, Place Jussieu, Tour 33, 75252
Paris Cedex 05, France
\and
Physique Corpusculaire et Cosmologie, College de
France,  11 pl. Marcelin Berthelot, F-75231 Paris Cedex 05, France
\and
Laboratoire de Physique Subatomique et de Cosmologie, 53 Avenue des
Martyrs, 38026 Grenoble Cedex, France
}

   \date{\today}

   \abstract{ We present a method designed to estimate the noise power
   spectrum in the time domain for CMB experiments.  The noise power
   spectrum is extracted from the time ordered data avoiding the
   contamination coming from sky signal and accounting the
   pixellisation of the signal and the projection of the noise when
   making intermediate sky projections.  This method is simple to
   implement and relies on Monte-Carlo simulations, it runs on a simple
   desk computer.  We also propose a trick for filtering data before
   making coadded maps in order to avoid ringing due to the presence of
   signal in the timelines.  These algorithms were succesfully tested
   on Archeops data.   }

   \maketitle

%________________________________________________________________

\section{Introduction}
Measuring the Cosmic Microwave Background anisotropies angular power
spectrum has proved to be a powerful cosmological tool giving direct
information on both the cosmological parameters and the primordial
Universe through the origin of the initial perturbations.

The usual method used nowadays for measuring the CMB temperature
fluctuations is to scan the sky with a detector (with a bolometer for
instance) that measures the temperature variations in a given beam.
This provides time data streams along with pointing directions that
are subsequently compressed into a sky map. In the general case, the
noise in the time streams is not white (due to atmospheric
contamination, spurious noises, electronic and detector temperature
fluctuations). The sky is also in general not regularly scanned so
that at the end, the noise in the maps is neither homogeneous nor
white.

This causes difficulties when trying to extract the signal power
spectrum from the maps as it has to be separated from the noise.  Fully
optimal methods finding the maximum likelihood solution for both the
map and the CMB power spectrum have been proposed such as
MADCAP~(\cite{borrill}) but they require the inversion of large
covariance matrices on parrallel supercomputers and are hard to
implement for the present generation of experiments that cover a large
portion of the sky and have a large number of pixels (WMAP, Archeops
and Planck in the near future).  Alternative methods that replace the
full inversion by a Monte-Carlo simulation of the noise angular power
spectrum have been proposed~(\cite{hivon,spice}).  The results obtained
with such methods are satisfactory and they are little time consuming
compared to the maximum likelihood methods.  The analysis process is
done in the following way in these Monte-Carlo methods :
\begin{enumerate}
\item{\bf noise power spectrum estimation in the data:\label{ptnoise}}
  it is very important not to be contaminated by signal at this stage.
  This is particularly difficult when dealing with large portions of
  the sky where galactic dust and clouds can be found. We propose in
  section~\ref{sectnoise} a method that provides an unbiased estimate
  of the noise Fourier power spectrum by correcting an initial guess
  via Monte-Carlo simulations.
\item{\bf Fast map making:\label{ptmap}} can be a simple coaddition of
  filtered timelines or a optimal iterative mapmaking that converges
  to the maximum likelihood solution for instance with a conjugate
  gradient method. This map making has to be very fast as it is
  repeated for each Monte-Carlo realisation. We propose in
  section~\ref{sectmap} a method for making simply coadded maps of
  filtered data avoiding the ringing due to the effect of the filtering
  on bright sources such as the Galactic plane.
\item{\bf $C_\ell$ estimation on the map:} This is done very easily
  with the Healpix package~(\cite{healpix}). The angular power
  spectrum obtained here is called {\em pseudo-$C_\ell$} as it differs
  from the true signal $C_\ell$ due to various effects: presence of
  noise, beam and time domain filtering effects on the signal,
  incomplete sky coverage that degrades the resolution in harmonic
  space. All these effects can be accounted for and corrected with the
  MASTER~(\cite{hivon}) or SpICE~(\cite{spice}) algorithms.
\end{enumerate}
In this article we propose methods for solving points \ref{ptnoise}
(in section~\ref{sectnoise}) and \ref{ptmap} (in
section~\ref{sectmap}). These methods have been sucessfully used for
the Archeops data analysis~(\cite{arch_cl,arch_param}). We used the
MASTER algorithm for going from {\em pseudo-$C_\ell$} to true $C_\ell$
and for estimating the error bars.

\section{Noise spectrum estimation\label{sectnoise}}
We present in this section the method we developped for determining
the noise Fourier power spectrum of a time data stream composed from
CMB fluctuations, Galactic dust and coloured gaussian and stationary
noise. The data is denoted as an $N$ elements vector $\vec{d}$ ($N$ is
the number of time samples) and is composed of noise $\vec{n}$ and sky
signal $\vec{s}$:
\begin{equation}\label{signoise}
\vec{d}=\vec{s}+\vec{n}
\end{equation}
The noise power spectrum is given by
$P(\nu)=\left<\vec{\tilde{n}}^*\vec{\tilde{n}}\right>$ where
$\tilde{~}$ is the Fourier transform and $\left<\right>$ the ensemble
average over many realisations. The point here is to avoid
contamination from signal (CMB and in a larger part Galactic dust)
that would tend to overestimate the estimated noise power spectrum and
therefore $\tilde{N}_\ell$ leading to a biased $C_\ell$ spectrum at
the end.

\subsection{Raw estimation of the noise spectrum}
The first raw estimation of the noise power can come from the data
itself (that is $\vec{n}\simeq\vec{d}$) as we are dealing with low
signal to noise experiments (at least in the time domain as 
the noise is reduced by redundancy in the map). This is a correct
approximation only in the case where the observed region of the sky
is free from any bright source.  In most recent experiments such as 
WMAP and Archeops (as well as in  Planck) the instrument covers a large portion
of the sky by making large circles that cross the Galactic plane twice
per rotation. There is therefore a very bright source in all the
portions of the time streams that prevent us from using the naive
approximation $\vec{n}\simeq\vec{d}$. 

A simple method to get rid of the signal contribution in the noise
power spectrum estimation is to project the data onto a map of the sky
portion observed.  We apply a prior highpass to the data to select the
frequancies we are interested in and to get a realistic map.  In such a
map, the noise level is reduced but the signal is almost unchanged.
Reading back this map with the scanning strategy leads to a timeline
where the signal to noise is greatly improved.  This latter timeline
can then be subtracted to the initial giving a noise estimate that is,
in principle, free from signal contamination.  In the matrix notation
that has become common in CMB analysis this corresponds to the
following operation:
\begin{equation} \label{firstapprox}
\vec{n}_{raw}=\vec{d}-A \left(A^T A\right)^{-1}A^T \vec{d}
\end{equation}
where $A$ is the pointing matrix.  The noise spectrum estimation is
then just obtained from the Fourier transform of $\vec{n}_{raw}$.

\subsection{Biasing effects}
If one rewrites equation~(\ref{firstapprox}) using the
definition~(\ref{signoise}), one gets:
\begin{eqnarray}
\begin{array}{ccccccc}
\vec{n}_{raw}&=&\vec{n}&-&\underbrace{A \left(A^T A\right)^{-1}A^T \vec{n}}
&+&\underbrace{\left(\vec{s}-A \left(A^T A\right)^{-1}A^T \vec{s}\right)}\\
&=&\vec{n}&-&\vec{r}_n&+&\vec{r}_s
\end{array}
\end{eqnarray}
where $\vec{r}_n$ and $\vec{r}_s$ can be interpreted in the following
way:
\begin{description}
\item{\bf residual noise ($\vec{r}_n$):} The number of
  time samples per pixel is finite (in Archeops and for a resolution
  of 15 arcminutes we have in average 25 samples per pixel with a very
  asymetric distribution) and therefore there remains some noise in
  the projected map. This noise in the map, once re--read with the
  scanning stategy is strongly correlated to the initial noise in the
  timeline. We therefore slightly underestimate the noise spectrum.
  Finally $\vec{r}_n$ increases if the pixel size is reduced. This
  residual noise is also discussed in~\cite{stomporetal}.
\item{\bf residual signal ($\vec{r}_s$):} Because of the
  pixellisation of the projected map, the high temporal frequencies of
  the signal (mainly dust a low galactic latitude or point sources)
  are badly subtracted from the initial time stream. This leads to an
  overestimation of the noise power spectrum. Finally $\vec{r}_s$
  decreases if the pixel size is reduced as it only originates from the large size of the pixels compared to the frequencies present in the data.
\end{description}
When computing the noise estimate power spectrum, one gets:
\begin{eqnarray}
P_{raw}(\nu)&=&\left<\left(\vec{\tilde n}-\vec{\tilde r}_n+
\vec{\tilde r}_s\right)^*
\left(\vec{\tilde n}-\vec{\tilde r}_n+\vec{\tilde r}_s\right)\right> \\
&=&
\left<\left(\vec{\tilde n}-\vec{\tilde r}_n\right)^*\left(\vec{\tilde n}
-\vec{\tilde r}_n\right)\right>
+\left<\vec{\tilde r}_s^*\vec{\tilde r}_s\right> \label{defspec}
\end{eqnarray}
because noise and signal are supposed uncorrelated. The first term can
be rewritten as $\left(1-\epsilon(\nu)\right)P(\nu)$. Simulations show
that $\epsilon(\nu)$ does not depend significantly on the noise power
spectrum in
the frequency range considered here ($\nu \ge 0.3$Hz), but essentially
on the scanning strategy. The second term is additive and depends on
the signal contained in the map and can be defined as
$\epsilon_s(\nu)$. One can therefore rewrite equation~(\ref{defspec})
as:
\begin{equation}\label{toinvert}
P_{raw}(\nu)=\left(1-\epsilon(\nu)\right)P(\nu)+\epsilon_s(\nu)
\end{equation}

In the case of the Archeops experiment these two effects are not
negligible at the resolutions we are interested in. As they cannot be
reduced simultaneously by increasing or decreasing the pixel size,
both have to be accounted for in order to correctly estimate the noise
power spectrum. In the following, we have chosen to use relatively
small pixel size (15 arcminutes) in order to reduce the contribution 
of $\epsilon_{s}$ to the bias as its determination, as will be seen
later, relies on calibration ans is model-dependent.

If one can determine the shape of both $\epsilon(\nu)$ and
$\epsilon_s(\nu)$, then debiasing the noise power is straighforward by
inverting equation~(\ref{toinvert}).

\begin{figure}[!t]
\resizebox{\hsize}{!}
{\includegraphics[clip]{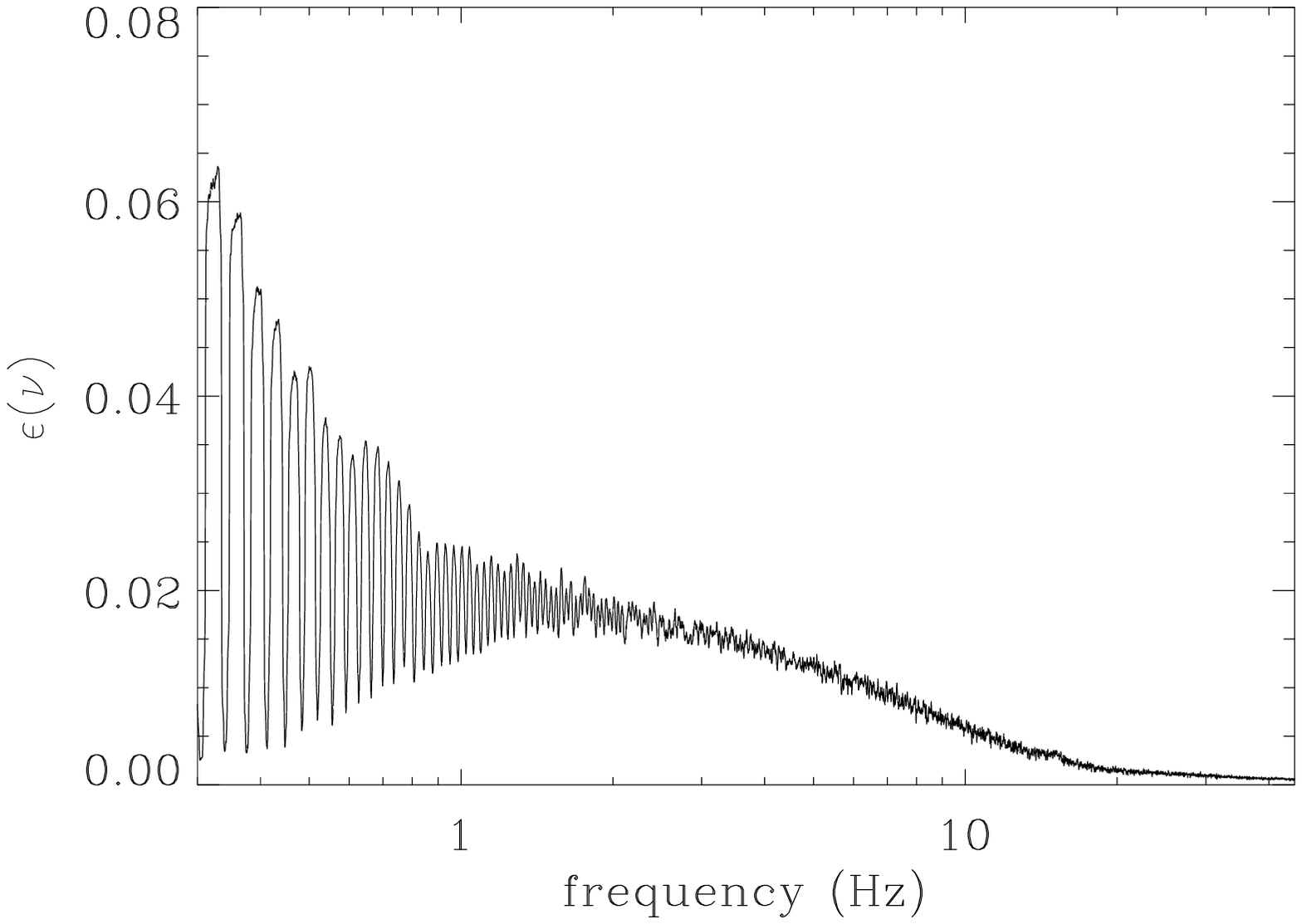}}
\caption{Noise bias $\epsilon(\nu)$ (multiplicative).}
\label{fig_noisebias_nu}
\resizebox{\hsize}{!}
{\includegraphics[clip]{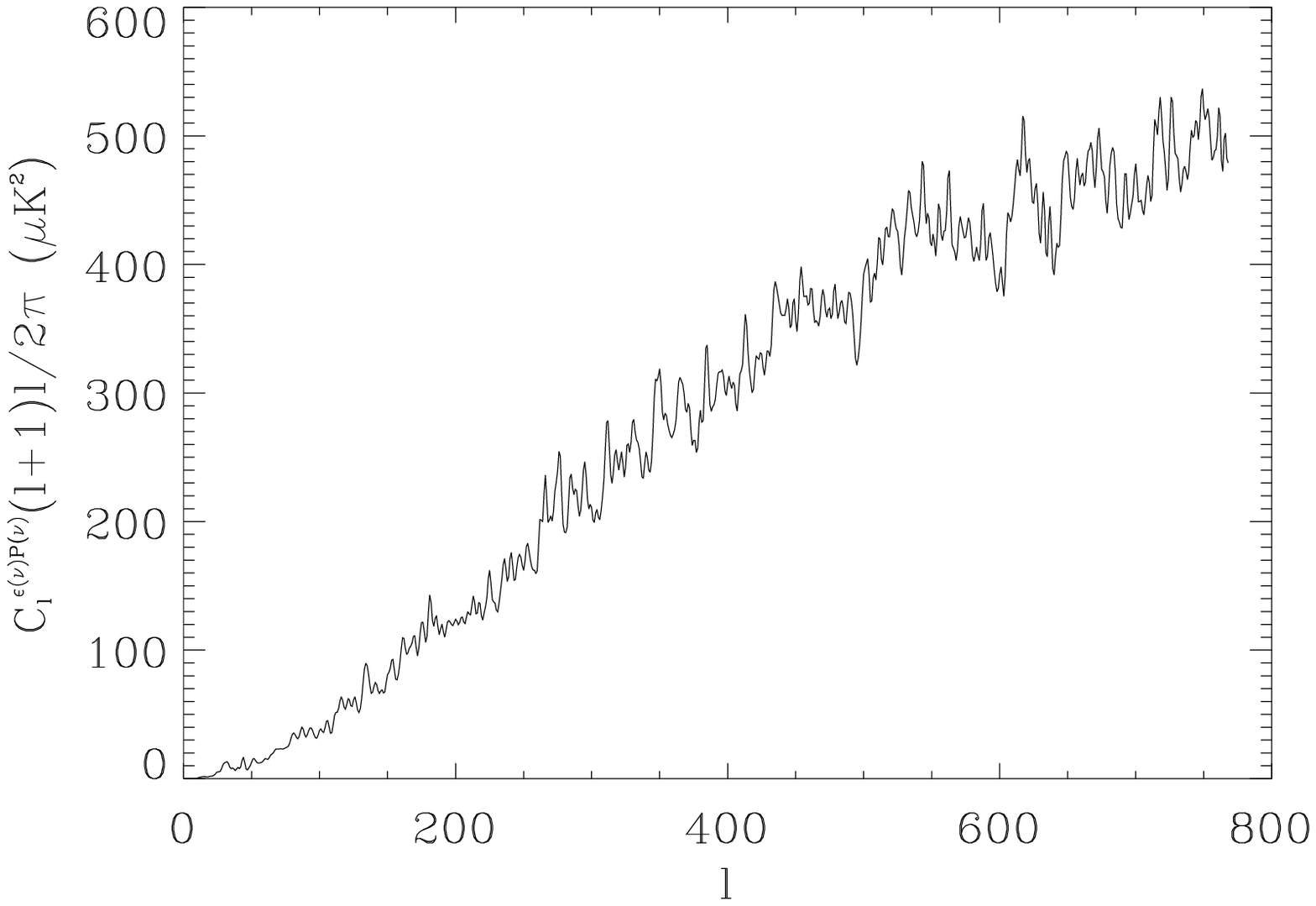}}
\caption{Power spectrum of the noise bias $\epsilon(\nu)\times P(\nu)$
in harmonic space.}
\label{fig_noisebias_nu_cl}
\end{figure}
\subsection{Estimating the noise effect}
The simplest way to estimate $\epsilon(\nu)$ is to make noise only
Monte-Carlo realizations of the data streams with spectrum
$P_{raw}$ (this assumes that $P_{raw}(\nu)$ is close enough to
$P$ so that $\epsilon$ is computed with enough accuracy altough the
algorithm could be iterated) and perform the operation of
equation~(\ref{firstapprox}) on each realisation. The average power
spectrum of all realizations is biased by $1-\epsilon(\nu)$ that can be
calculated by taking the ratio of the input power spectrum to the
output.  The result obtained is shown in Fig.~\ref{fig_noisebias_nu}.
Large oscillations can be seen at lower frequencies with peaks at
harmonics of the spinning frquency.  This is not surprising as we
expect the noise to project onto the sky at these frequencies.  It is
therefore at these particular frequencies that the noise bias is the
largest, leading to large values of $\epsilon(\nu)$.  The noise bias
transformed into harmonic power spectrum using the Archeops pointing
strategy is shown in Fig.~\ref{fig_noisebias_nu_cl} showing that this
effect should not be neglected as it amounts to a significant fraction
of the error bars (shown in black in Fig.~\ref{fig:ringing}).

\begin{figure}[!t]
\resizebox{\hsize}{!}
{\includegraphics[clip]{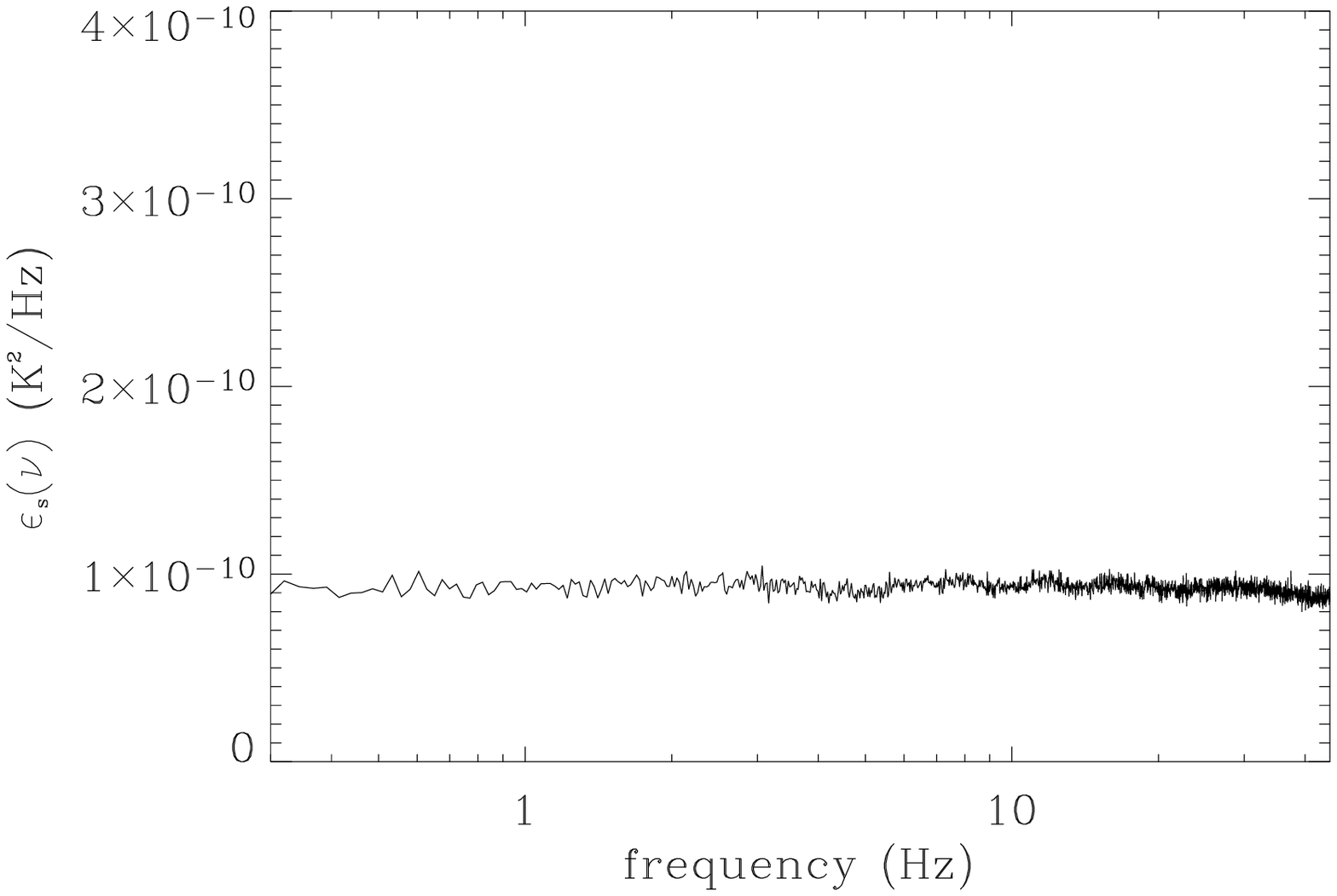}}
\caption{Signal Bias $\epsilon_s(\nu)$ (additive).}
\label{fig_signalbias_nu}
\resizebox{\hsize}{!}
{\includegraphics[clip]{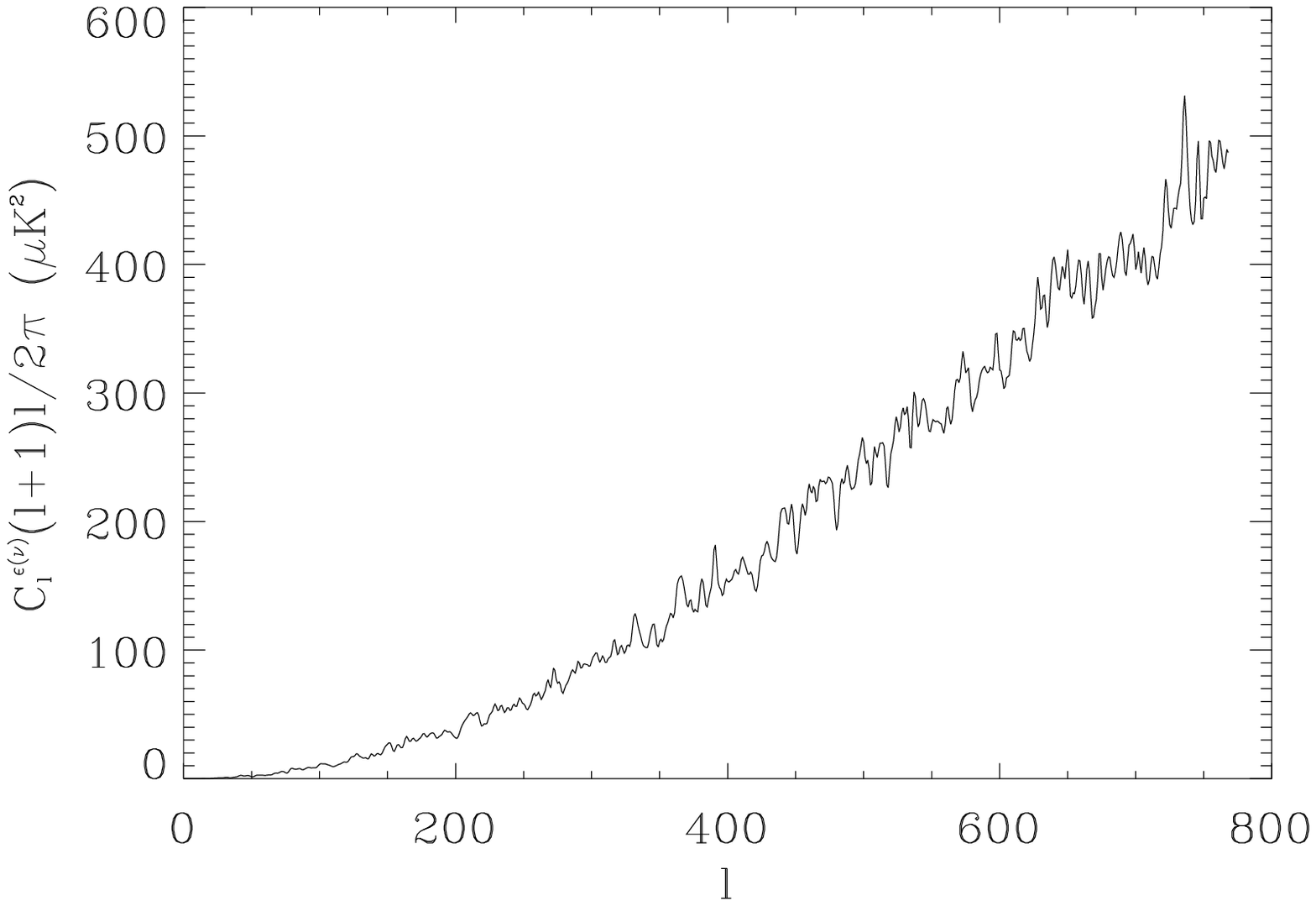}}
\caption{Power spectrum of the signal bias $\epsilon_s(\nu)$ in harmonic space.}
\label{fig_signalbias_nu_cl}
\end{figure}

\subsection{Estimating the signal effect}
The signal bias estimation does not require averaging over realizations
as it depends only on the sky signal, dominated by Galactic dust.  The
estimation is simply done by adding realistic Galactic dust signal on a
noise only timeline (with spectrum $P_{raw}(\nu)$).  The realistic
simulated Galactic dust signal is obtained from \cite{SFD} dust maps
filtered with the corresponding time constants, beam and electronic
filter.  The noise+signal timeline is then deglitched and filtered the
same way as for the real data.  We then remove the initial noise before
applying the operation of equation~(\ref{firstapprox}).  The galactic
bias $\epsilon_s(\nu)$ is the Fourier power spectrum of the residual
signal.  The signal bias $\epsilon_s(\nu)$ is shown in
Fig.~\ref{fig_signalbias_nu} in Fourier space and in
Fig.~\ref{fig_signalbias_nu_cl}.  Again, the power spectrum bias
amounts to a significant fraction of the error bars, showing that the
signal bias should not be neglected.  The level of galactic bias highly
depends on the level of Galactic signal in the data and therefore the
calibration.  The biggest uncertainty in the estimation of
$\epsilon_s(\nu)$ therefore arises from the uncertainty in the
calibration typically around 10$\%$ (in temperature).  It is finally
relatively small compared to the signal as it only applies to the
signal residual that is already a few percent of the whole signal.

\begin{figure*}[!t]
\resizebox{\hsize}{!}
{\includegraphics[clip]{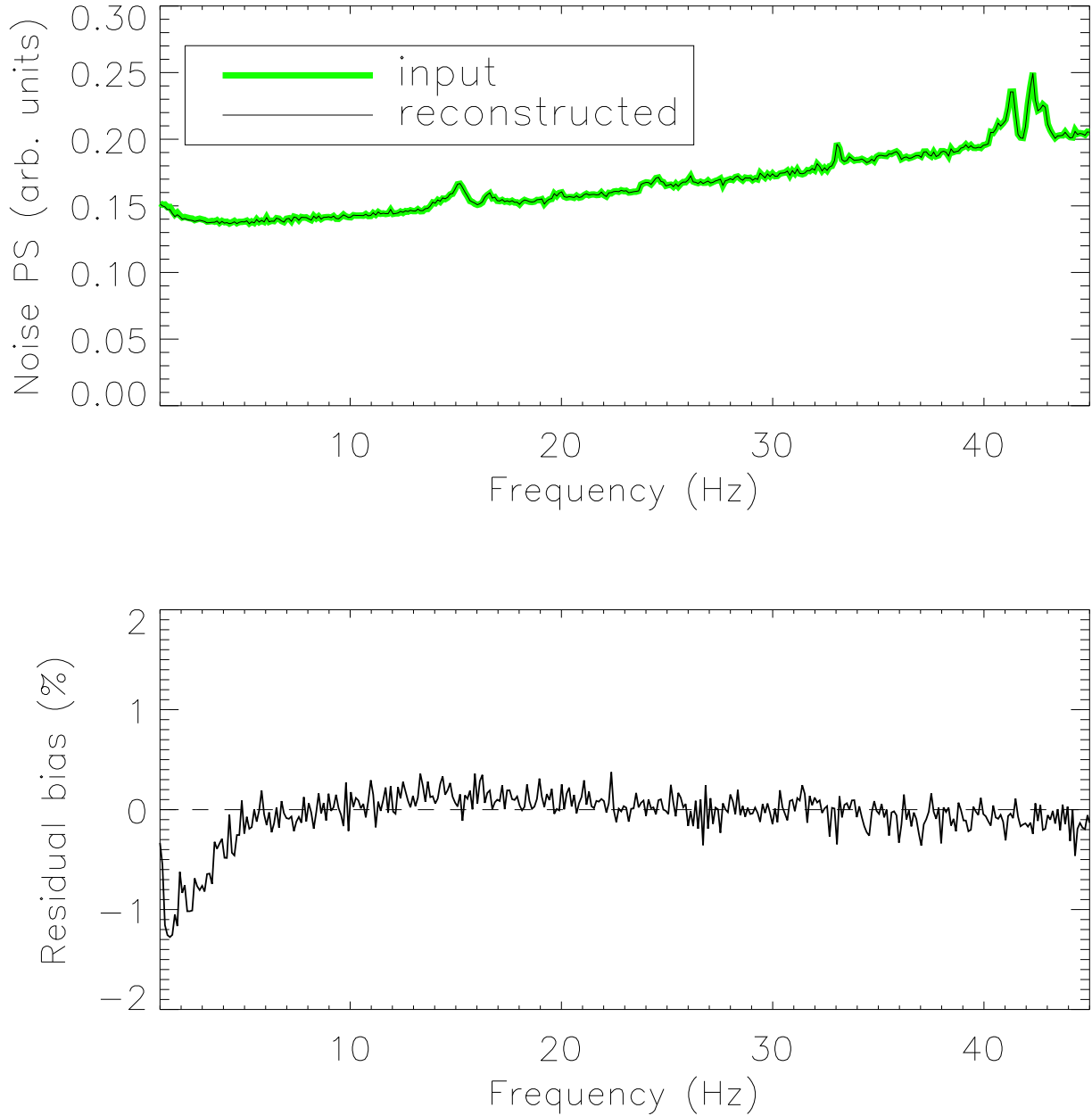}}
\caption{Results of the noise spectrum reconstruction obtained with realistic 
realizations. The top panel shows the initial noise spectrum in green
and the average reconstruction over the 60 realisations in black. The
lower panel shows the relative residual bias which is much smaller
than 2 percent at all frequencies.}
\label{fig_res_spec}
\end{figure*}

\subsection{Accuracy on the reconstructed noise spectrum}
After debiasing of both effects, we obtain a noise spectrum that
should, in principle reflect the initial noise spectrum very
accurately. In order to estimate this effect, we performed 60
simulations of the debiasing procedure based on the same inital noise
spectrum (a realistic Archeops noise spectrum). The result is shown in
Fig.~\ref{fig_res_spec}. The noise reconstruction is better than two
percents at all frequencies which is sufficient for accurate noise
spectrum subtraction with MASTER in the case studied here.

\section{Making maps with filtered data : filtering without ringing\label{sectmap}}
Making coadded maps using heavily filtered data (in the Archeops case only
frequencies between 0.3 and 45 Hz are kept) is a simple and accurate alternative
to optimal mapmaking when the knee frequency of the noise is low
enough to prevent the filtering from removing too much signal in the
low frequency part. Even when this is the case, one has to make sure
that the filtering (obtained by setting low Fourier modes to zero) has
the expected effect and does not pollute the data.

Spurious effects from filtering do arise if there is strong signal in
the data. A good example of such a strong signal is the Galactic plane
crossings that occur twice per rotation with Archeops. As the galactic
peak is essentially non stationary the filtering creates ringing
signal around the galactic plane in the maps. The structures created
are not only located just around the galactic plane (where they are
the strongest) but all over the map at a level comparable to the CMB
anisotropies that are searched for. We therefore need a trick to
remove these ringing effects, that is, filtering without ringing.

The usual trick used to reduce ringing effects on filtered data is
not to use too sharp filters. We used a filter with a sine shape going
from 0 to 1 in 0.06 Hz. We obtained this way good reduction of the
ringing effect but we wanted to improve further the mapmaking.

\begin{figure}[!t]
\resizebox{\hsize}{!}
{\includegraphics[clip]{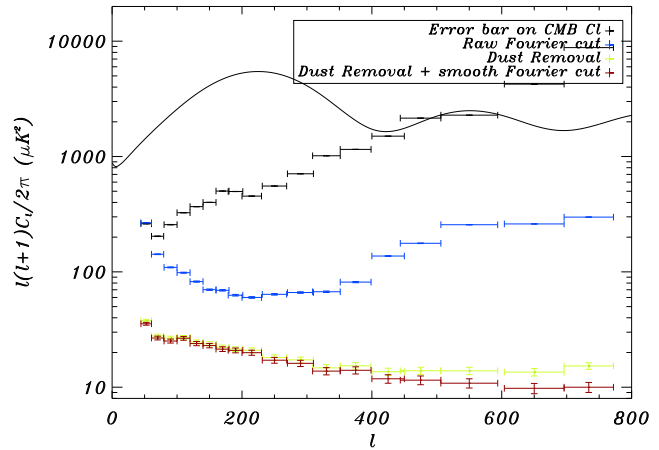}}
\caption{The error bars ont the power spectrum estimation are shown shown 
in black. The excess power spectrum contribution coming from the
ringing effect on the Galactic signal is shown in blue for a raw
Fourier cut on the inital data.. The effect of the dust removal by
subtracting our estimate of the dust signal is shown in
green. Adding the smooth sine shape filter instead of a raw Fourier
cut leads to the red residuals.}
\label{fig:ringing}
\end{figure}

As the ringing arises from the presence of the Galaxy crossing in the
data, removing the Galactic signal should reduce drastically the
effect. A first step is to remove simulated galactic signal from
\cite{SFD} but the agreement between the simulated maps and the true Galactic 
signal is not perfect and there remains high peaks in the data. This
method also introduces external data into the experimental data, a procedure
which has to be avoided if possible. We therefore decided to remove
an estimate of the Galactic signal done with the experimental data itself. This is done in
the following way:
\begin{itemize}
\item We first fit the data with a set of low frequency functions (sinc 
functions multiplied by a gaussian~(\cite{bourrachot})) not taking into account data
located in bright regions of the Galaxy (using a mask made from
\cite{SFD} galactic dust maps). The low frequency template obtained is 
then free from Galactic contamination.
\item We then remove this low frequency template from the initial data
and obtain a timeline containing mostly high frequency noise and
Galactic signal and no low frequency stripes.
\item This timeline is then simply coadded to obtain a map. We then set
to zero the pixels out of the Galactic mask used before so that the
map contains Galactic signal in the region known to be contaminated
and zero elsewhere (in particular, in the region where the CMB
analysis is to be performed).
\item This map is read with the scanning strategy and the resulting timeline 
is subtracted from the initial data. The resulting timeline is stricly
identical to the initial one in the CMB region and differs in the
Galactic region just by the estimate of the Galactic signal.
\item This last timeline is then filtered using a sine shape filter between 
0.3 and 45 Hz and coadded to make the map that will be used for the CMB
analysis.
\end{itemize}
This method allows a very efficient removal of the galactic signal in
the contaminated region without changing anything in the clean region
where one wants to perform the CMB analysis. The efficiency of the
ringing effect removal is shown in Fig.~\ref{fig:ringing}.  The
ringing coming from the Galactic signal is efficiently reduced to a
level much smaller than the CMB anisotropies and the error bars on the
power spectrum estimation.

\section{Conclusions}
We have developped a method that allows an accurate estimate of the
noise power spectrum for CMB experiments accounting for small but
significant effects such as the noise reprojection and the
pixellisation.  This method relies on Monte-Carlo simulations and does
not require specific platforms to be implemented.  It is designed to
serve as an input to CMB power spectrum estimation that generally
Requires a precise estimate of the noise properties.  This method has
been succesfully applied in the frame of the Archeops data
analysis~(\cite{arch_cl,arch_param}) but can be extended to any non
interferometric experiment.

%________________________________________________________________
\begin{acknowledgements}
The authors are grateful to the whole Archeops collaboration for
uncountable stimulating discussions which have made this work
possible. We wish to thank in particular M.~Douspis,
J.-F.~Mac\'{\i}as-P\'erez and F.-X.~D\'esert for their contributions.
\end{acknowledgements}

\end{document}